\documentclass[11pt,draftcls,onecolumn]{IEEEtran}

\usepackage{epsf}
\usepackage{graphics}
\usepackage{times}
\usepackage{graphicx}
\usepackage{amsmath}
\usepackage{amssymb}
\usepackage{epsfig}
\usepackage{subfigure}
\usepackage{array}
\usepackage{float}

\begin{document}

\title{Fast Correlation Computation Method for Matching Pursuit Algorithms in Compressed Sensing}

\author{Kee-Hoon~Kim,~Hosung~Park,~Seokbeom Hong,~Jong-Seon~No,~and~Habong~Chung
\thanks{K.-H. Kim, H. Park, S. Hong, and J.-S. No are with the
Department of Electrical Engineering and Computer Science, INMC, Seoul
National University, Seoul, 151-744, Korea (phone: +82-2-880-8437,
fax: +82-2-880-8222, email: kkh@ccl.snu.ac.kr, lovepk98@snu.ac.kr, fousbyus@ccl.snu.ac.kr, jsno@snu.ac.kr).}
\thanks{H. Chung is with the School of Electronics and Electrical Engineering,
Hong-Ik University, Seoul 121-791, Korea (e-mail: habchung@hongik.ac.kr).}
\thanks{This work was supported by the National Research Foundation of Korea (NRF) grant funded by the Korea government (MEST) (No. 2012-0000186).}
}

\maketitle

\begin{abstract}
There have been many matching pursuit algorithms (MPAs) which handle the sparse signal recovery problem a.k.a. compressed sensing (CS). In the MPAs, the correlation computation step has a dominant computational complexity. In this letter, we propose a new fast correlation computation method when we use some classes of partial unitary matrices as the sensing matrix. Those partial unitary matrices include partial Fourier matrices and partial Hadamard matrices which are popular sensing matrices. The proposed correlation computation method can be applied to almost all MPAs without causing any degradation of their recovery performance. And, for most practical parameters, the proposed method can reduce the computational complexity of the MPAs substantially.

\end{abstract}


\section{Introduction}

Compressed sensing (CS) is a novel sampling technique, where one can recover sparse signals from the undersampled
measurements \cite{Candes}. In a typical CS problem, the goal is to exactly reconstruct the $N \times 1$ $K$-\textit{sparse signal vector} $x$ based on the $M \times 1$ \textit{measurement vector} $y$. By $K$-sparse we mean that there are at most $K$ nonzero elements in $x$. The vectors $x$ and $y$ are linearly related to each other as
\begin{equation}\label{eq:CSbasic}
y = \Phi x + \eta,
\end{equation}
where $\Phi$ is the $M\times N$ \textit{sensing matrix} and $\eta$ is the $M \times 1$ noise vector. And the relation of $K$, $M$, and $N$ is generally $K<M \ll N$.

For the sensing matrix $\Phi$, partial Fourier matrices and partial Hadamard matrices are popular sensing matrices, where we mean that the partial matrix is constructed by some $M$ rows of $N \times N$ original matrix $A$. In other words, $\Phi = S_\Omega A$, where $S_\Omega$ is the $M \times N$ row selection matrix consisting of $M$ rows (indices from some index set $\Omega$) of $N \times N$ identity matrix $I$.

Firstly, the partial Fourier matrix is frequently used because of its good recovery performance, fast implementation using the fast Fourier transform (FFT), and applicability to practical signals. The examples include channel estimation in communication systems \cite{Bajwa} and magnetic resonance imaging (MRI) \cite{Lustig}. For the partial Fourier matrix, the index set $\Omega$ can be constructed randomly or based on the cyclic difference set \cite{Yu}.

Secondly, some recent researches showed that well-designed deterministic sensing matrices based on linear block codes have better performance and less complexity for signal recovery compared to random sensing matrices \cite{Calderbank}, \cite{Hong}. It is well known that a sensing matrix whose columns are bipolar-presented codewords of a binary linear block code can be viewed as a partial Hadamard matrix. And we can exploit the efficiency of the fast Hadamard transform (FHT).

To recover $x$ in (\ref{eq:CSbasic}), matching pursuit algorithms (MPAs) find a sparse estimation of the signal $x$ from $y$ in a greedy fashion. It works iteratively by choosing the component that has the highest correlation with the current residual. Examples include the orthogonal matching pursuit (OMP) \cite{Tropp} and its modified versions such as the compressive sampling matching pursuit (CoSaMP) \cite{Needell}, the regularized OMP (ROMP) \cite{Needell2}, the subspace pursuit (SP) \cite{Dai}, and the backtracking-based matching pursuit (BB MP) \cite{Huang}. For instance, we summarize the OMP which is the most basic algorithm among the MPAs. The steps marked by $\lozenge$ are the common steps to the MPAs.

\rule{.95\linewidth}{0.2mm}

\textbf{Algorithm 1.1 Conventional OMP recovery algorithm}

\rule{.95\linewidth}{0.2mm}
\begin{enumerate}
  \item Initialize : $r_0=y$, $\Lambda_0={\o}$, $t=1$. $\lozenge$
  \vspace{3pt}
  \item Correlation computation : $h_{t-1} = \Phi^H r_{t-1}$. $\lozenge$
  \vspace{3pt}
  \item Identification : $\lambda_t=\mathrm{arg~max}_{j=1,...,N}|h_{t-1}(j)|.$
  \vspace{3pt}
  \item Augment the index set : $\Lambda_t = \Lambda_{t-1} \cup \{\lambda_t \}$.
  \vspace{3pt}
  \item Construct $\Phi_t$ : $\Phi_t = \Phi S_{\Lambda_t}^T$. $\lozenge$
  \vspace{3pt}
  \item Least squares : $x_t = (\Phi_t^H\Phi_t)^{-1}\Phi_t^H y$.
  \vspace{3pt}
  \item Update current residual : $a_t = \Phi_t x_t$, $r_{t} = y - a_t$. $\lozenge$
  \vspace{3pt}
  \item $t = t+1$, return to 2) if the halting criterion is not triggered. $\lozenge$
\end{enumerate}

\rule{.95\linewidth}{0.2mm}

In Algorithm 1.1, performing $h_{t-1} = \Phi^H r_{t-1}$ in 2) can be viewed as computing the correlations between the current residual $r_{t-1}$ and the columns of $\Phi$. And we denote $h_{t-1}$ as the correlation vector at the $t$-th iteration. The whole computational complexity of the OMP is dominated by the correlation computation step and so are the other MPAs'.

In this letter, we propose a new fast correlation computation method which can be applied to almost all MPAs including OMP, CoSaMP, ROMP, SP, and BB MP. The recovery performances of the MPAs applied by the proposed method are exactly the same as those of the original MPAs. And, for most practical parameters, the proposed method can reduce the computational complexity of the MPAs substantially. The proposed method can operate only when the sensing matrix is the partial unitary matrix satisfying the following two constraints :
\begin{enumerate}
\item Every element of the unitary matrix $U$ has the magnitude $1/\sqrt{N}$.
\item The set $\{ \sqrt{N}u_1, \sqrt{N}u_2, \cdots, \sqrt{N}u_N \}$, where $u_n$ is the $n$-th column of $U$, is closed under element-wise multiplication $\circ$.
\end{enumerate}

At a glance, the above constraints seem to be too strict, however, the Fourier matrix and the Hadamard matrix are two kinds of the unitary matrices with these constraints. Therefore, the proposed method is meaningful and it can be widely adopted in CS.

\section{A New Fast Correlation Computation Method for MPAs}

In this section, we describe the proposed fast correlation computation method for general MPAs. The MPAs have the common steps marked by $\lozenge$ in Algorithm 1.1 and we derive the fast correlation computation method based on only those steps. In the following derivation, $U$ is the unitary matrix satisfying the two constraints and the sensing matrix is $\Phi = S_\Omega U$. For simplicity, we handle not the $t$-th iteration but the $(t+1)$-th iteration.

Using the steps 5) and 7) in Algorithm 1.1, the correlation computation step 2) at the $(t+1)$-th iteration $h_t = \Phi^H r_t$ can be rewritten as
\begin{align}\label{eq:corr_com}
h_t     =& U^H S_\Omega^T r_{t}\nonumber\\
        =& U^H S_\Omega^T (y - a_t)\nonumber\\
        =& U^H S_\Omega^T y - U^H S_\Omega^T a_t\nonumber\\
        =& h_0 - U^H S_\Omega^T S_\Omega U S_{\Lambda_t}^T x_t,
\end{align}
where $h_0 = \Phi^H r_0 = U^H S_\Omega^T y$.

In (\ref{eq:corr_com}), $S_{\Lambda_t}^T x_t$ can be represented as
\begin{equation}\label{eq:superposition}
S_{\Lambda_t}^T x_t = \sum_{\tau=1}^{|\Lambda_t|} x_t(\tau) e_{\Lambda_t(\tau)},
\end{equation}
where $e_{\Lambda_t(\tau)}$ is the ${\Lambda_t(\tau)}$-th column of $I$, $|\Lambda_t|$ is the cardinality of the index set $\Lambda_t$, and $x_t(\tau)$ is the $\tau$-th element of $x_t$.
By using (\ref{eq:corr_com}) and (\ref{eq:superposition}), we obtain
\begin{align}\label{eq:superposition2}
h_t =& h_0 - U^H S_\Omega^T S_\Omega U \sum_{\tau=1}^{|\Lambda_t|} x_t(\tau) e_{\Lambda_t(\tau)} \nonumber\\
    =& h_0 - \sum_{\tau=1}^{|\Lambda_t|} x_t(\tau) U^H S_\Omega^T S_\Omega U e_{\Lambda_t(\tau)}.
\end{align}

Without loss of generality, we assume the first column of $U$ is $(1/\sqrt{N}, 1/\sqrt{N}, \cdots, 1\sqrt{N})^T$.
And (\ref{eq:superposition2}) can be rewritten as
\begin{equation}\label{eq:superposition3}
h_t = h_0 - \sum_{\tau=1}^{|\Lambda_t|} x_t(\tau) U^H S_\Omega^T S_\Omega D_{\Lambda_t(\tau)} U e_1,
\end{equation}
where $D_{\Lambda_t(\tau)} = \sqrt{N}\cdot\mathrm{diag}(u_{\Lambda_t(\tau)})$ and $u_{\Lambda_t(\tau)}$ is the ${\Lambda_t(\tau)}$-th column of $U$. Because the matrix $S_\Omega^T S_\Omega$ is the diagonal matrix, $S_\Omega^T S_\Omega D_{\Lambda_t(\tau)} = D_{\Lambda_t(\tau)} S_\Omega^T S_\Omega$ and (\ref{eq:superposition3}) can be rewritten as
\begin{equation}
h_t     = h_0 - \sum_{\tau=1}^{|\Lambda_t|} x_t(\tau) U^H D_{\Lambda_t(\tau)} S_\Omega^T S_\Omega U e_1.
\end{equation}

We denote $P_{\Lambda_t(\tau)} = U^H D_{\Lambda_t(\tau)} U$ and thus $P_{\Lambda_t(\tau)} U^H = U^H D_{\Lambda_t(\tau)}$. Consequently, the correlation computation at the $(t+1)$-th iteration can be expressed as
\begin{align}\label{eq:kernelsum}
h_{t}     =& h_0 - \sum_{\tau=1}^{|\Lambda_{t}|} x_t(\tau) P_{\Lambda_{t}(\tau)} U^H S_\Omega^T S_\Omega U e_1 \nonumber\\
        =& h_0 - \sum_{\tau=1}^{|\Lambda_{t}|} x_t(\tau) P_{\Lambda_{t}(\tau)} c,
\end{align}
where $c = U^H S_\Omega^T S_\Omega U e_1$ which is called the \textit{correlation kernel vector}. Note that the correlation kernel vector $c$ is independent to the sparse signal vector $x$ and thus can be stored in advance.
The matrix $P_{\Lambda_t(\tau)}$ in (\ref{eq:kernelsum}) is a permutation matrix according to the following theorem. The permutation matrix can be performed with negligible computational complexity because of its structure.

\vspace{2.5pt}

\textbf{Theorem 2-1} : $P_{\Lambda_t(\tau)} = U^H D_{\Lambda_t(\tau)} U$ is a permutation matrix (i.e., a square binary matrix that has exactly one element 1 in each row and each column and $0$s elsewhere) if the unitary matrix $U$ is under the two constraints.

\vspace{2.5pt}
\textit{Proof of Theorem 2-1} : $U^H D_{\Lambda_t(\tau)} U$ can be expressed as
\begin{align}\label{eq:appendix1}
U^H D_{\Lambda_t(\tau)} U =& \left( u_1~\cdots~u_N \right)^H \cdot \sqrt{N}\cdot \mathrm{diag}(u_{\Lambda_t(\tau)}) \cdot \left( u_1~\cdots~u_{N} \right) \nonumber\\
                          =& \frac{1}{\sqrt{N}} \cdot \left( u_1~\cdots~u_N \right)^H \cdot \left( \sqrt{N}u_{\Lambda_t(\tau)}\circ\sqrt{N}u_1~\cdots~\sqrt{N}u_{\Lambda_t(\tau)}\circ\sqrt{N}u_N \right).
\end{align}

$\sqrt{N}u_{\Lambda_t(\tau)}\circ\sqrt{N}u_n$, $n=1,\cdots,N$, are distinct column vectors because their elements are nonzero by the first constraint of $U$. And each vector belongs to the set $\{ \sqrt{N}u_1, \sqrt{N}u_2, \cdots, \sqrt{N}u_N \}$ because of the second constraint of $U$. Therefore, (\ref{eq:appendix1}) can be rewritten as
\begin{align}
U^H D_{\Lambda_t(\tau)} U =& \frac{1}{\sqrt{N}} \cdot \left( u_1~\cdots~u_N \right)^H \cdot \left( \sqrt{N}u_1~\cdots~\sqrt{N}u_N \right) \cdot P_{\Lambda_t(\tau)}\nonumber\\
                          =& \left( u_1~\cdots~u_N \right)^H \cdot \left( u_1~\cdots~u_{N} \right) \cdot P_{\Lambda_t(\tau)}\nonumber\\
                          =& I P_{\Lambda_t(\tau)} = P_{\Lambda_t(\tau)},
\end{align}
where $P_{\Lambda_t(\tau)}$ is the permutation matrix which is determined by $\Lambda_t(\tau)$ and the structure of $U$. $\square$

To sum it up, the correlation computation at the $t$-th iteration (i.e., computing $h_{t-1}$) can be performed by $|\Lambda_{t-1}|$ subtractions of properly scaled and permutated versions of the correlation kernel vector $c$ to the initial correlation vector $h_0$.


\section{Fast OMP Recovery Algorithm}

In this section, we apply the fast correlation computation method to the conventional OMP. And we discuss the complexity of the proposed OMP algorithm applied by the proposed method. Applying the proposed correlation computation method to other MPAs is straightforward and entirely analogous with this section.

\subsection{Fast Correlation Computation for the OMP}

The proposed correlation computation method (\ref{eq:kernelsum}) for a general MPA can be easily converted for the OMP as
\begin{equation}\label{eq:kernelsumOMP}
h_{t-1}= \begin{cases} \Phi^H r_0,  & t=1 \\
                      h_0 - \sum_{\tau=1}^{t-1} x_{t-1}(\tau) P_{\lambda_{\tau}} c, & t>1.
            \end{cases}
\end{equation}
And the proposed OMP recovery algorithm can be given by simply replacing the correlation computation step 2) in Algorithm 1.1 with (\ref{eq:kernelsumOMP}).

Note that the proposed OMP algorithm is actually identical to the conventional OMP algorithm. The only difference is the computation method and thus the proposed OMP guarantees the same recovery performance compared to the conventional OMP.

\subsection{Complexity Analysis}

In this subsection, we investigate the computational complexity of the proposed OMP algorithm in the cases of using the partial Fourier matrix and the partial Hadamard matrix. Firstly, for each matrix, we will discuss the properness of the proposed algorithm in terms of the storage requirements for the permutation matrices. Secondly, we compare the computational complexity of the proposed OMP algorithm to that of the conventional OMP algorithm. For the exact comparison, we consider the number of flops of each algorithm. And we regard one complex multiplication as $6$ flops and one complex addition as $2$ flops.

We remark that the proposed OMP performs the $(t-1)$ subtractions of properly scaled and permutated versions of the correlation kernel vector at the $t$-th iteration to compute the correlation vector in the second equation in (\ref{eq:kernelsumOMP}). We consider the case when $N$ is a power of two, which is used very often in signal processing. But, the proposed method can be used for any $N$.

\subsubsection{Storage Requirements Using the Partial Fourier Matrix} When we use the partial Fourier matrix as the sensing matrix, from the discrete Fourier transform (DFT) properties, $P_{\lambda_{\tau}}$ is the matrix which cyclically shifts $c$ when $P_{\lambda_{\tau}}$ is multiplied with the vector $c$ from the left. Therefore, the storage requirements for the permutation matrices can be negligible.

\subsubsection{Computational Complexity Using the Partial Fourier Matrix}
It is well known that the correlation computation at each iteration in the conventional OMP can be implemented by the $N$-point FFT. The $N$-point FFT requires $5N\log_2 N$ flops.

For the proposed OMP, in (\ref{eq:kernelsumOMP}), the first iteration ($t=1$) can be implemented by one FFT. And, when $t>1$, the correlation vector is computed by using the correlation kernel vector. Exploiting the conjugate symmetric property of the correlation kernel vector using the partial Fourier matrix as the sensing matrix, performing the second equation in (\ref{eq:kernelsumOMP}) has the cost of $(N/2)(t-1)$ complex multiplications, $2(N/2)(t-1)$ real additions, and $N(t-1)$ complex additions at the $t$-th iteration. Aggregately, the proposed OMP requires $6N(t-1)$ flops at the $t$-th iteration.

\begin{table}
\caption{Comparison Between the Proposed OMP and the Conventional OMP (\# of flops at the $t$-th iteration)}
\label{Table:Fourier}
\centering
\begin{tabular}{|c|c|c|}
\hline
\multicolumn{ 3}{|c|}{ $\Phi$ : Partial Fourier matrix} \\
\hline
     t      &        $=1$ &        $>1$ \\
\hline
Conventional OMP &     $5N\log_2 N$ &     $5N\log_2 N$ \\
\hline
Proposed OMP &     $5N\log_2 N$ &    $6N(t-1)$ \\
\hline
\multicolumn{ 3}{c}{} \\
\hline
\multicolumn{ 3}{|c|}{$\Phi$ : Partial Hadamard matrix} \\
\hline
  t      &        $=1$ &        $>1$ \\
\hline
Conventional OMP  &     $2N\log_2 N$ &     $2N\log_2 N$ \\
\hline
Proposed OMP  &     $2N\log_2 N$ &    $2N(t-1)$ \\
\hline
\end{tabular}
\end{table}

\subsubsection{Storage Requirements Using the Partial Hadamard Matrix}
The storage requirements of the proposed OMP algorithm with partial Hadamard matrix are also favorable. There is no need to store the entire $N$ permutation matrices. If we store only the $\log_2 N$ permutation matrices, the permutation matrix $P_{\lambda_{\tau}}$ for any $\lambda_{\tau}$ can be easily performed by sequentially applying some matrices among the stored $\log_2 N$ permutation matrices. It is easily induced from the properties of the Hadamard matrix.

\subsubsection{Computational Complexity Using the Partial Hadamard Matrix}
It is well known that the correlation computation at each iteration in the conventional OMP can be implemented by the $N$-point FHT. The $N$-point FHT requires $N \log_2 N$ complex additions (i.e., $2N\log_2 N$ flops).

For the proposed OMP, in (\ref{eq:kernelsumOMP}), the first iteration ($t=1$) can be implemented by one FHT. And, when $t>1$, the correlation vector is computed by using the correlation kernel vector. Because the correlation kernel vector consists of only a small number of values compared to $N$ using the partial Hadamard matrix as the sensing matrix, performing the second equation in (\ref{eq:kernelsumOMP}) has approximately the cost of $N(t-1)$ complex additions. Table I summarizes this subsection. 

\section{Numerical Analysis}

Here we present some numerical results characterizing the performance of the proposed OMP algorithm compared to the conventional OMP algorithm. The results were produced using the partial Fourier matrices and the partial Hadamard matrices with practical and various sizes.
And we plot the computational complexities for $1\leq t \leq 13$, which is reasonable for given $M$ and $N$.

\begin{figure}[H]
\centering
\includegraphics[width=.9\linewidth]{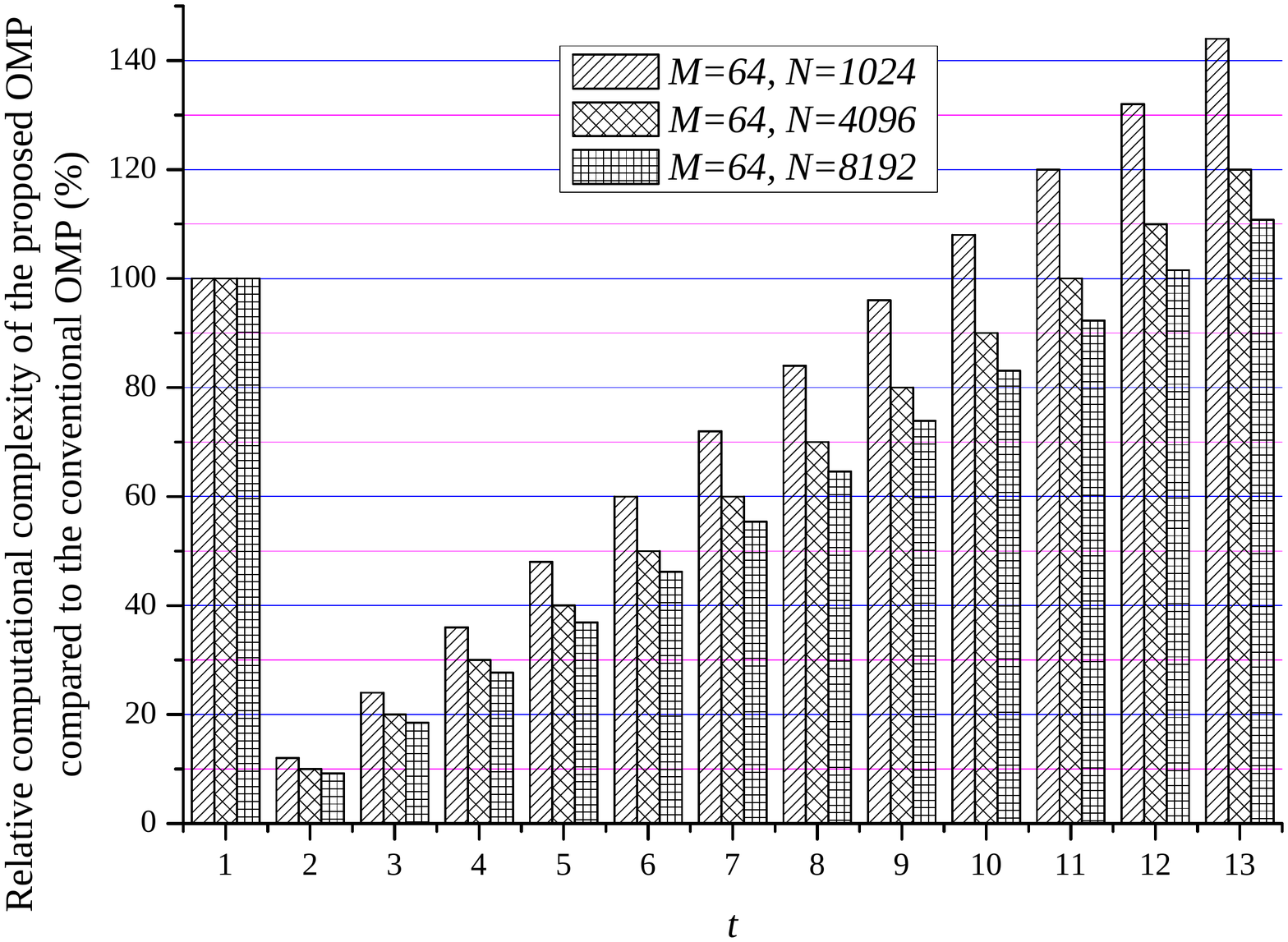}
\caption{Relative computational complexity of the proposed OMP compared to the conventional OMP at the $t$-th iteration when the partial Fourier matrices are used.}
\label{fig:OMP_3N}
\end{figure}


Fig. \ref{fig:OMP_3N} shows the relative computational complexity of the proposed OMP compared to the conventional OMP when the partial Fourier matrices are used. Because the computational complexity of the proposed OMP algorithm at the $t$-th iteration is proportional to $t-1$, there is a excessive point and thus adaptive strategy is needed. For instance, for $M=64$ and $N=4096$, $t=11$ is the excessive point and the conventional OMP can be used from the $12$-th iteration. Consequently, the proposed OMP algorithm has a benefit to reduce the computational complexity substantially. Especially, for large $N$, the proposed OMP has a good benefit.


\begin{figure}[H]
\centering
\includegraphics[width=.9\linewidth]{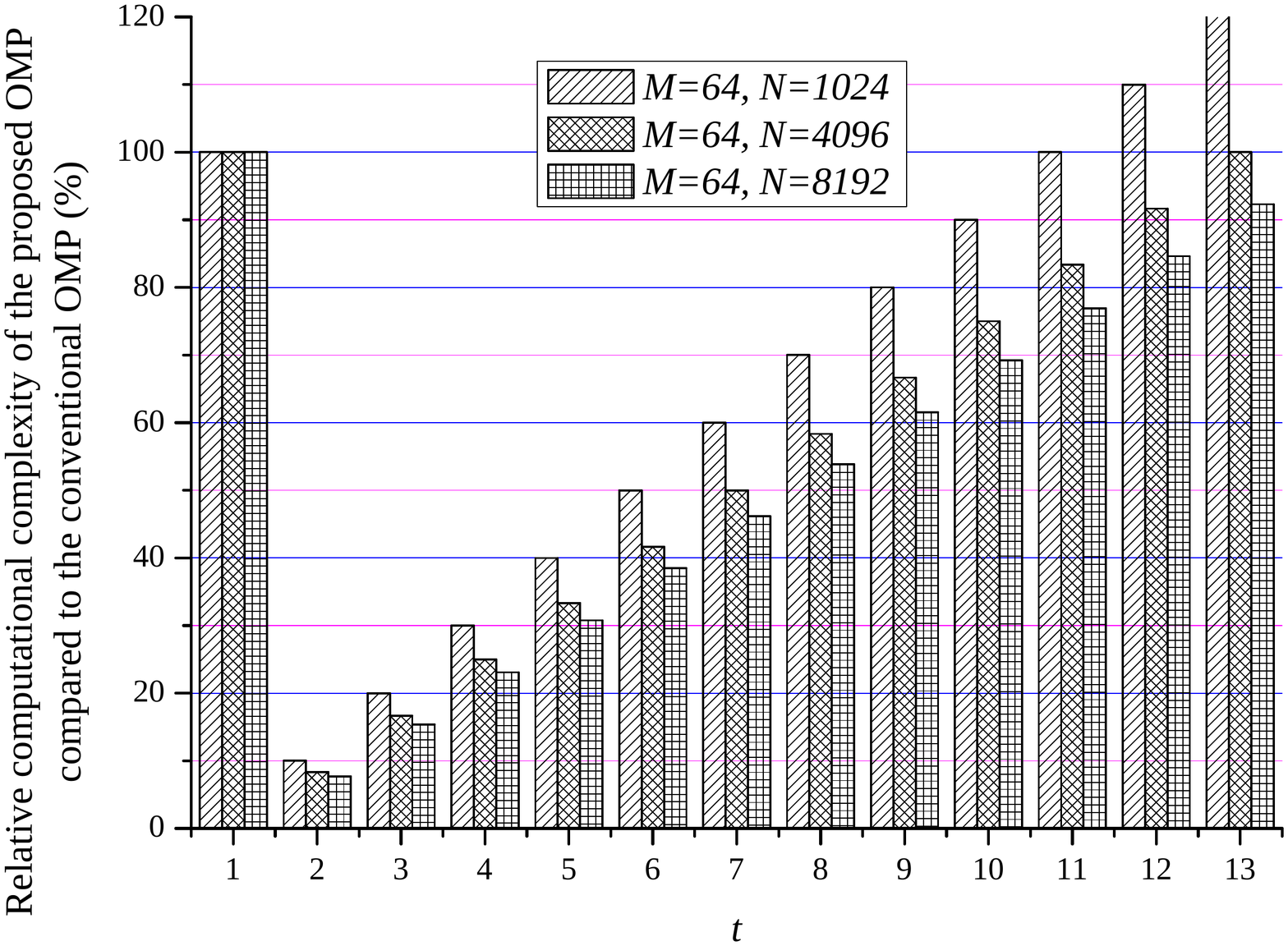}
\caption{Relative computational complexity of the proposed OMP compared to the conventional OMP at the $t$-th iteration when the partial Hadamard matrices are used.}
\label{fig:OMP_3N_HDM}
\end{figure}

Fig. \ref{fig:OMP_3N_HDM} shows the relative computational complexity of the proposed OMP compared to the conventional OMP when the partial Hadamard matrices are used. Like the case of using the partial Fourier matrices in Fig. \ref{fig:OMP_3N}, the proposed OMP algorithm has a benefit to reduce the computational complexity substantially.

\begin{figure}[H]
\centering
\includegraphics[width=.9\linewidth]{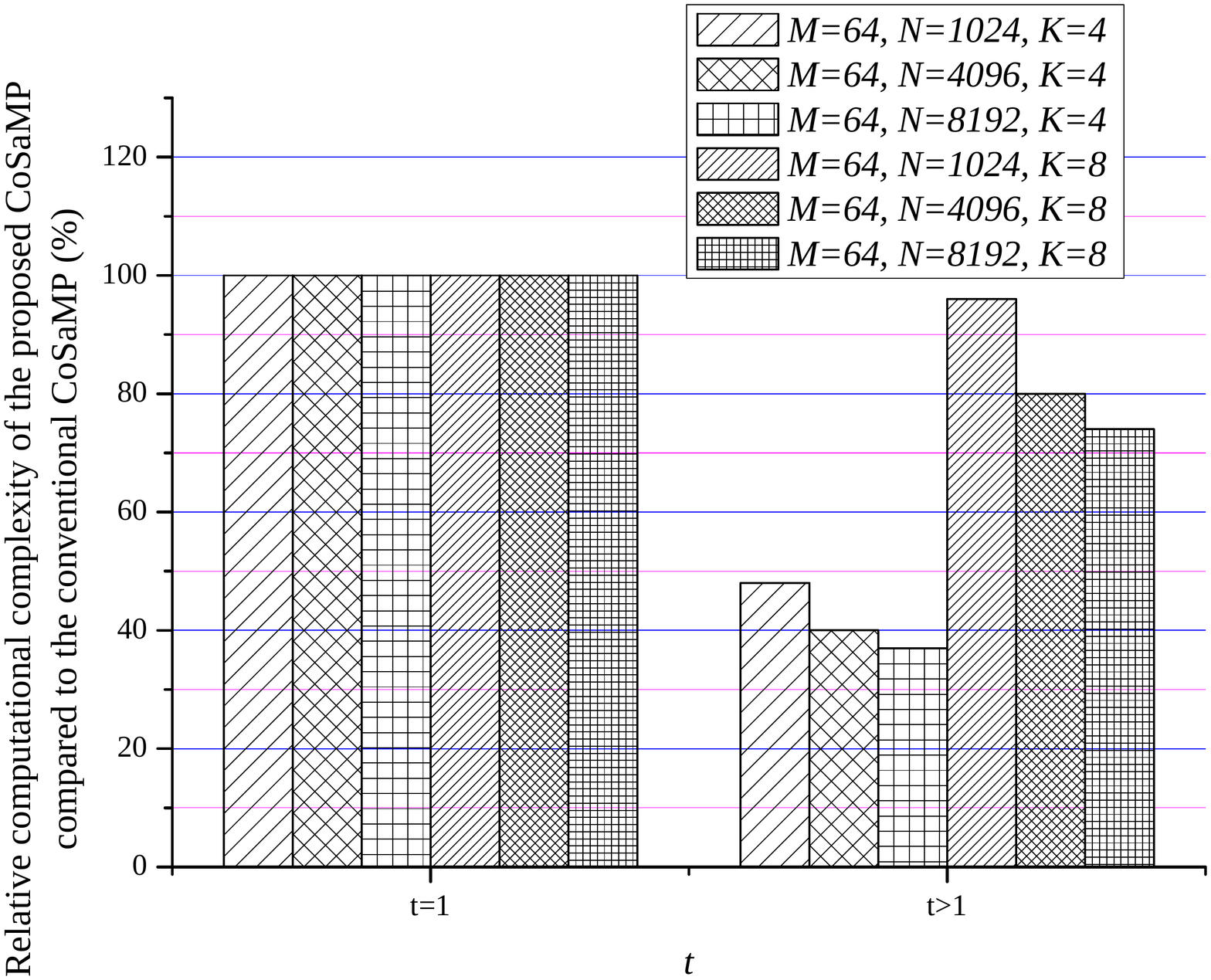}
\caption{Relative computational complexity of the proposed CoSaMP compared to the conventional CoSaMP at the $t$-th iteration when the partial Fourier matrices are used.}
\label{fig:CSMP}
\end{figure}

Besides the proposed OMP, we present the numerical result when the proposed correlation computation method is applied to the CoSaMP~\cite{Needell}. Due to lack of space, we leave out the detailed description of the proposed CoSaMP. Fig. \ref{fig:CSMP} shows the relative computational complexity of the proposed CoSaMP compared to the conventional CoSaMP at the $t$-th iteration. We use the partial Fourier matrices as the sensing matrix. Different to the proposed OMP, the proposed CoSaMP requires the same computational cost at each iteration except when $t=1$. Especially, the proposed CoSaMP algorithm has a good benefit for small $K$ and large $N$. For instance, when $M=64$, $N=8192$, and $K=4$, the proposed CoSaMP requires only the $37\%$ computational cost compared to the conventional CoSaMP.

\end{document}